# Scanning tunneling spectroscopy of the surface states of Dirac fermions in thermoelectrics based on bismuth telluride

L N Lukyanova, I V Makarenko, O A Usov and P A Dementev

Ioffe Institute, Russian Academy of Sciences
26 Politekhnicheskaya, St Petersburg 194021, Russian Federation

E-mail: lidia.lukyanova@mail.ioffe.ru

**Abstract.** Morphology of the interlayer van der Waals surface and differential tunneling conductance in p-$Bi_{2-x}Sb_xTe_{3-y}Se_y$ solid solutions were studied by scanning tunneling microscopy and spectroscopy in dependence on compositions. Topological characteristics of the Dirac fermion surface states were determined. It is shown that the thermoelectric power factor and the material parameter enhance with the shift of the Dirac point to the top of the valence band with increasing of atomic substitution in these thermoelectrics. Correlation between topological characteristics, power factor and material parameter was found. A growth contribution of the surface states is determined by an increase of the Fermi velocity for large atomic substitutions of Bi at $x>1.5$ and small substitutions in the Te sublattice ($y=0.06$). In compositions with smaller substitutions at $x = (1-1.3)$ and $y=(0.06-0.09)$, similar effect of the surface states is determined by raise of the surface concentration of charge carriers.

## 1. Introduction

Solid solutions $(Bi, Sb)_2(Te, Se)_3$ at optimal composition and charge carrier concentration for temperature range from 120 to 500 K are known to be high-performance thermoelectrics that are widely used in thermoelectric heating and cooling [1-3]. Nanostructuring of chalcogenide thermoelectrics allows to enhance the figure-of-merit Z in comparison with bulk materials due to decrease in the thermal conductivity, increase in the power factor and the material parameter [4-7].

Novel opportunities for the development of innovative chalcogenide materials of a new generation are associated with studies of the topological properties of the Dirac fermion surface states in bismuth telluride and its solid solutions belonging to the class of topological insulators. In these materials, topological state appears as the result of electronic band inversion due to strong spin-orbit interaction, thus the bulk becomes an insulator [8, 9] and the surface demonstrates an unusual metallic properties specific for Dirac fermions with linear dispersion and helical spin texture that ensures the lack of electron backscattering on nonmagnetic defects. The anomalous properties of the topological surface electronic states are now intensively applied in photonics [10] and optoelectronics [11, 12], special attention is also paid to their application in thermoelectricity [13-16].

The work is devoted to study of the surface states of Dirac fermions in p-type $(Bi, Sb)_2(Te, Se)_3$ thermoelectrics with substitution of atoms in the Bi and Te sublattices by scanning tunneling microscopy (STM) and spectroscopy (STS). At present, STM/STS methods are used to obtain the local characteristics of the surface electronic states of Dirac fermions [17-20], along with angle-resolved photoelectron spectroscopy (ARPES) [17, 21-23] and transport properties [24-26].

STM/STS permit to determine the peculiarities of morphology and tunneling differential conductance that define the topological surface states, namely the Dirac point energy $E_D$, the top of the valence band $E_V$, the bottom of the conduction band $E_C$, the energy gap $E_g$, and the surface concentration of Dirac fermions $n_s$ in dependence on the of solid solution composition.

Joint analysis of topological surface state parameters of Dirac fermions and bulk properties of thermoelectrics [21-22] allows to establish a correlation between the STM/STS results and the Seebeck coefficient, the power factor and the material parameter, which are proportional to the density-of-states effective mass, the charge carrier mobility and the figure-of-merit Z.

## 2. Crystal structure and samples for studies

Van der Waals thermoelectrics based on bismuth telluride are a rhombohedral crystal structure with a $R\bar{3}m(D_{3d}^5)$ space group. The parameters of the primitive rhombohedral cell for $Bi_2Te_3$ are $a_R$ = 1.0476 nm, the angle between the basis vectors is 24.166°, the parameters of the corresponding hexagonal cell $a$ and $c$ are 0.4385 and 3.0487 nm, respectively.

The crystalline structure of bismuth telluride consists of plane anisotropic layers, each containing five atomic planes forming quintuple, which are separated by van der Waals gaps. The chemical bonds in the layers are mainly covalent with some ionicity. The atomic layers of Te and Bi in the quintuple alternate in the sequence (-Te (1) -Bi-Te (2) -Bi-Te (1)-) as shown in figure 1.

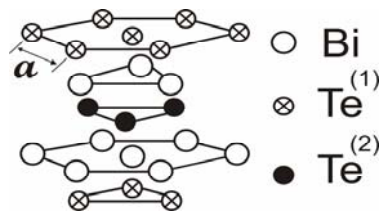

Figure 1. Quintuple of bismuth telluride

The ingots of p-$Bi_2Te_3$ and its based solid solutions with substitutions of atoms Bi → Sb and Te → Se were grown by the method of vertical zone alignment using precision stabilization of temperature at the crystallization front for obtaining homogeneous thermoelectrics [27]. The excess of Te atoms in comparison with stoichiometric composition were introduced in the materials for optimization of charge carrier concentration.

Table1. Seebeck coefficient S, power parameter $S^2\sigma$ and material parameter $(m/m_0)^{3/2}\mu_0$ in bulk thermoelectrics at T = 300 K*

| N | Composition | S, $\mu V\ K^{-1}$ | $S^2\sigma$, $10^{-6}$ W cm$^{-1}$ K$^{-2}$ | $(m/m_0)^{3/2}\mu_0$, $10^3$ cm$^2$ V$^{-1}$ s$^{-1}$ |
|---|---|---|---|---|
| 1 | p-Bi$_2$Te$_3$ | 252 | 32 | 190 |
| | p-Bi$_{2-x}$Sb$_x$Te$_{3-y}$Se$_y$, | | | |
| 2 | x=1, y=0.06 | 258 | 28 | 290 |
| 3 | x=1.15, y=0.06 | 251 | 29.5 | 361 |
| 4 | x=1.2, y=0.09 | 280 | 22 | 328 |
| 5 | x=1.3, y=0.09 | 273 | 30 | 478 |
| 6 | x=1.55, y=0 | 174 | 47 | 580 |
| 7 | x=1.6, y=0.06 | 168 | 47 | 567 |

* $\sigma$ is the electrical conductivity, $m/m_0$ is the density-of-states effective mass, and $\mu_0$ is the mobility of the charge carriers, obtained with taking into account degeneracy.

Anisotropic layered samples were cut from single-crystal grains of bulk ingots along the van der Waals (0001) interlayer planes oriented along the growth axis, which is perpendicular to the axis of third order *c*. These samples were cleaved along the interlayer planes to obtain the samples for STM/STS measurements.

The parameter $(m/m_0)^{3/2}\mu_0$ (Table 1) was determined from studies of thermoelectric and galvanomagnetic properties in the framework of the parabolic model of the energy spectrum with isotropic scattering mechanism taking into account its dependence on solid solution composition [28-32].

## 3. STM and STS methods

STM/STS studies of thermoelectrics based on bismuth and antimony chalcogenides were carried out using high–vacuum GPI–300 microscope developed at Prokhorov General Physics Institute (Moscow) with a new vacuum module developed and manufactured in Ioffe Institute, St. Petersburg. Before the experiments the several upper layers of sample were removed in STM chamber (~ 1.5x10$^{-7}$ Pa) by adhesive tape to obtain a clean surface. STM/STS tips were fabricated from tungsten wire with diameter of 260 μm, etched in 2M solution of NaOH to achieve the apex radii ~ 15 nm. Then the tips were heated in the preliminary chamber (~ 3.5 x10$^{-6}$ Pa) at 600° ÷ 700°C followed by a several hours cleaning by Ar$^+$ ion bombardment.

The structure of the surfaces (0001) was registered in the constant–current mode with feedback turned on. After that, in the regime of a constant distance between the probe and the surface under investigation (feedback turned off) the dependences of both $I_t = f_1(U_t)$ and $dI_t/dU_t = f_2(U_t)$ at frequencies up to 7 kHz were recorded at preselected points with averaging of up to 100 experimental curves to increase the signal/noise ratio.

**4. Surface morphology**

The morphology of the interlayer van der Waals surface (0001) was studied by STM in the p-$Bi_2Te_3$ and the p-$Bi_{2-x}Sb_xTe_{3-y}Se_y$ solid solutions with different substitutions of atoms in the Bi and Te sublattices (Table 1).

Investigation of the surface (0001) morphology with atomic resolution in the constant current mode at $I_t = (0.2 - 0.3)$ nA, $U_t = (250 - 800)$ mV showed the high quality of the hexagonal closed packed surface structure, despite the presence of dark and white regions, which nevertheless do not distort atomic periodicity of the structures [33–35]. The clearest image of Te atoms were obtained for $Bi_2Te_3$ (figure 2a). The typical surfaces (0001) of solid solutions with Bi → Sb atom substitutions at x> 1 have more diffuse images (figure 2b, 2c). Besides, the longwave modulation along the height profile of the (0001) surface is observed in figure 2b, 2c that is related to the substitutions of Bi → Sb, Te → Se atoms in solid solutions.

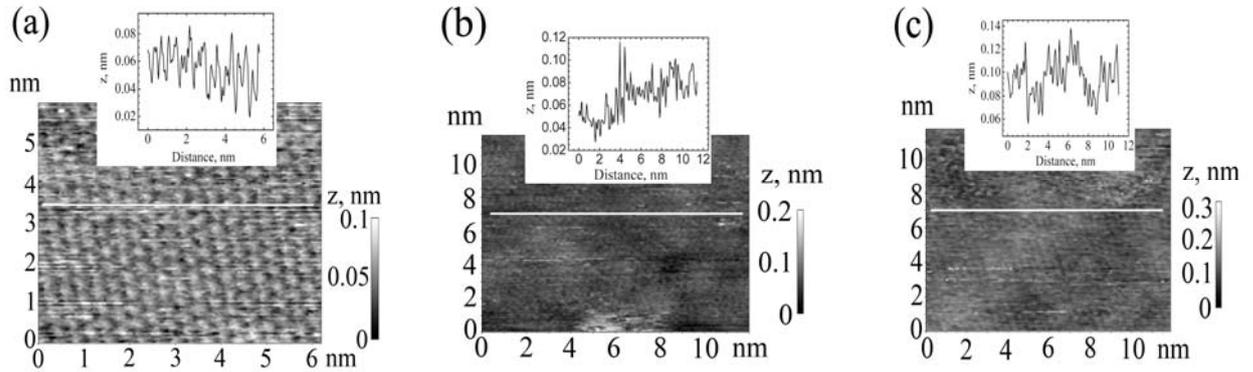

Figure 2. The morphology of the (0001) surface of the p-$Bi_2Te_3$ (a) and the solid solutions of $Bi_{2-x}Sb_xTe_{3-y}Se_y$ (x=1, y=0.06) (b), (x=1.6, y=0.06) (c), obtained in the regime $I_t$, nA, $U_t$, mV: a – 0.2 and 800, b – 0.2 and 350, c – 0.3 and 250. The inset of figures 2a – 2c shows the surface profiles taken along the white lines, respectively.

From the analysis of the surface profiles, the hexagonal lattice parameters were estimated in the range a = 0.40–0.48 nm that is in agreement with [36]. Corrugation of surface Te(1) atoms is about 0.06 nm in p-$Bi_2Te_3$ and about 0.1 nm in the solid solutions (figure 2a – 2c). Such height difference is attributed to the distortion of the surface electronic states that arise by the substitution of Bi → Sb, Te → Se atoms in the solid solutions [34, 35] or the formation of structural defects [33].

The Fast Fourier Transform (FFT) pictures of the (0001) surface in p-$Bi_2Te_3$ (figure 2a) and solid solutions demonstrate the intensity spectra of two-dimensional reciprocal space with reciprocal lattice points centered at the Γ point of the Brillouin zone (figure 3). The features in the intensity of the spectral components observed in the vicinity of the Γ point are supposed to be specific for the interference of quasiparticle excitations of surface electrons on defects (figures 3, a – d) known as Friedel oscillations in metals [37], topological insulators $Bi_{1.5}Sb_{0.5}Te_{1.7}Se_{1.3}$ [34] and $(Bi_{1-x}Sb_x)_2Te_3$ [35]. Besides, more intense second-order spectral components were observed in the solid solution with an increase of Se content at x = 1.2 and y = 0.09 as compared with x = 1.6, y = 0.06 (figures 3b, 3d).

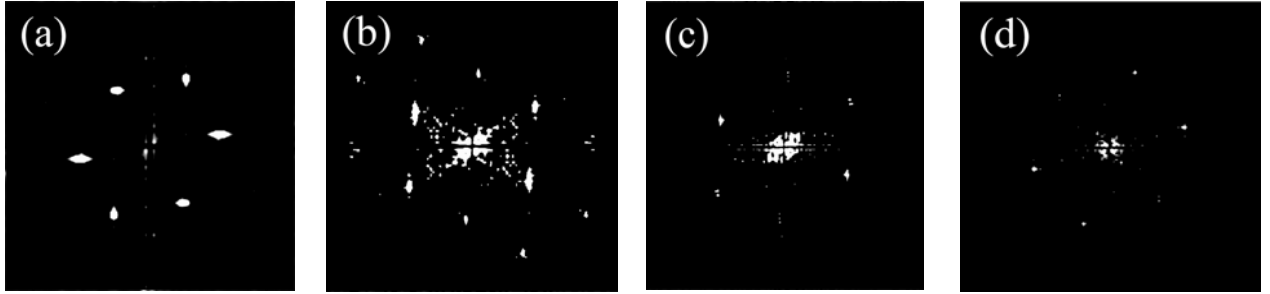

Figure 3. FFT of the (0001) surface images in: $Bi_2Te_3$ – (a), $Bi_{2-x}Sb_xTe_{3-y}Se_y$: (b) – x= 1.2, y=0.09, (c) – x= 1.55, y=0, (d) – x = 1.6, y=0.06

The FFT analysis of the surface images (0001) with atomic resolution was used to determine the average value of the hexagonal lattice constant (figure 3a – 3c). The average period for $Bi_2Te_3$ is about 0.45 nm and decreases in the solid solutions $Bi_{0.7}Sb_{1.3}Te_{2.91}Se_{0.09}$ and in $Bi_{0.45}Sb_{1.55}Te_3$ to 0.41 nm. These values of the lattice constant turned out to be less than the values determined from the height profile on the surface (0001), due to STM drift during image recording.

## 5. STS spectra

The differential tunneling conductance $dI_t/dU_t$ in dependence on voltage $U_t$ was measured in the p-$Bi_2Te_3$ and the p-$Bi_{2-x}Sb_xTe_{3-y}Se_y$ solid solutions on the (0001) surface by STS at room temperature. Figure 4 shows the differential conductance $dI_t/dU_t$ on voltage $U_t$, for the p-$Bi_{2-x}Sb_xTe_3$ (x=1.55) solid solution. The Dirac point energy $E_D$ was determined from the intersection of the tangent to the linear part of $dI_t/dU_t$ and the $U_t$ axis (figure 4) [38]. Slope changes along function $f(U_t) = dI_t/dU_t$ allow to locate the energy position of the top of the valence $E_V$ and the bottom of the conduction $E_C$ band. The $E_V$ and $E_C$ energy positions were refined using the second derivative $dI^2_t/dU^2_t$.

Similarly, the differential conductance $dI_t/dU_t = f(U_t)$ for p-$Bi_2Te_3$ and solid solutions of the different compositions were treated (figure 5). In solid solutions of p-$Bi_{2-x}Sb_xTe_{3-y}Se_y$, the position of the Dirac point $E_D$ shifts towards the top of the valence band in comparison with p-$Bi_2Te_3$, but in the investigated materials the Dirac point remains inside the bulk valence band (figure 5).

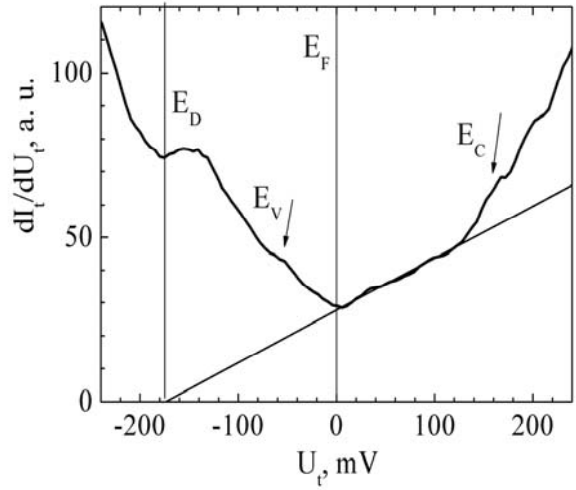

Figure 4. Differential conductance $dI_t/dU_t$ versus $U_t$ in the p-$Bi_{2-x}Sb_xTe_3$ (x= 1.55) solid solution. $E_D$ is the Dirac point energy. $E_V$ and $E_C$ are energies of the valence and the conduction band edges, respectively. $E_F$ is the position of Fermi level.

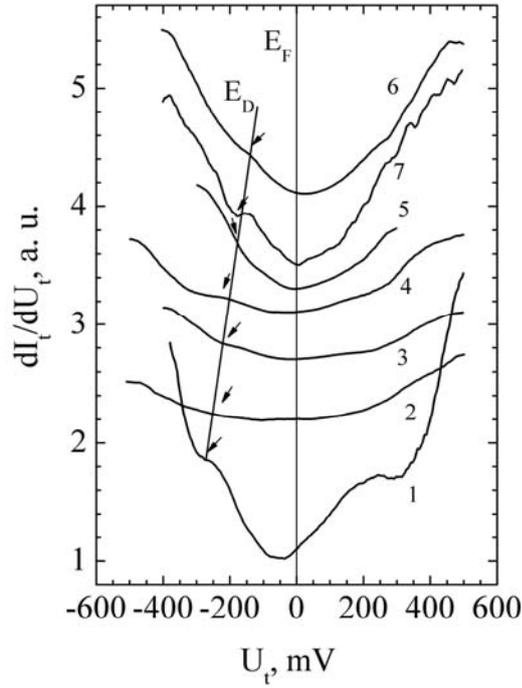

Figure 5. Normalized differential conductance $[dI_t/dU_t]_{norm} = dI_t/dU_t/[dI_t/dU_t]_{min}$ versus $U_t$ (each spectrum is vertically shifted for clarity). The line and arrows indicate the position of Dirac point $E_D$.
p-$Bi_2Te_3$ – *1*, p-$Bi_{2-x}Sb_xTe_{3-y}Se_y$: (x= 1, y=0.06) – *2*, (x= 1.15, y=0.06) – *3*, (x= 1.2, y=0.09) – *4*, (x= 1.3, y=0.09) – *5*, (x= 1.55, y=0) – *6*, (x= 1.6, y=0.06) – *7*.

The $E_D$ shift increases with growth of the atomic substitution in the Bi sublattice. The $E_D$ positions in p-$Bi_2Te_3$ and the solid solution p-$Bi_{0.45}Sb_{1.55}Te_3$ obtained in this study (figure 6) are in good agreement with [39]. The substitution of Te → Se atoms leads to an additional shift of the $E_D$ position towards the top of the valence band even with insignificant amounts of Se atoms in p-$Bi_{2-x}Sb_xTe_{3-y}Se_y$ solid solutions at y = 0.06 and 0.09 as compared with solid solutions of p-$Bi_{2-x}Sb_xTe_3$ at x = (1–1.3) (figure 6) [39]. The effect of Se atoms was also observed from the analysis of FFT picture in figure 3b, where the intensive second-order spectral components appear for composition with increased Se content to y = 0.09.

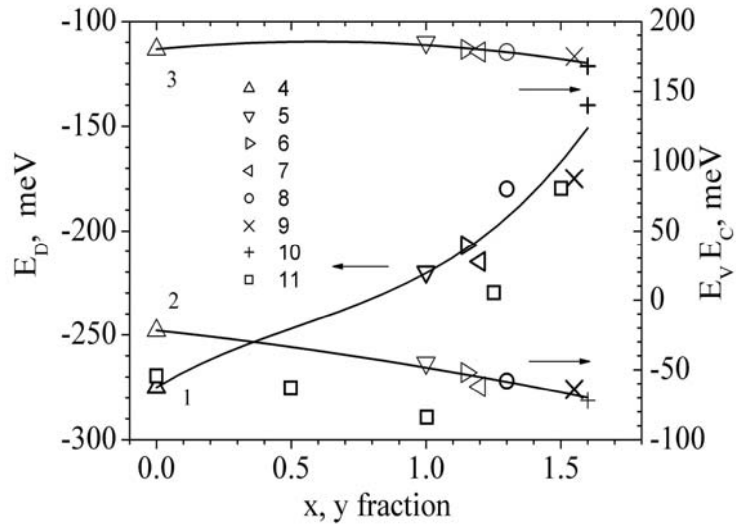

Figure 6. Position of the Dirac point $E_D$ (1), the top of the valence band $E_v$ (2) and the bottom of the conduction band $E_c$ (3) in dependence on the composition of p-$Bi_2Te_3$ and solid solutions. p-$Bi_2Te_3$ – *4*, p-$Bi_{2-x}Sb_xTe_{3-y}Se_y$: *5* – (x= 1, y=0.06), *6* – (x= 1.15, y=0.06), *7* – (x= 1.2, y=0.09), *8* – (x= 1.3, y=0.09), *9* – (x= 1.55, y=0), *10* – (x= 1.6, y=0.06), *11* – [39].

The growth of the atomic substitutions in solid solutions results in displacement of the edges of the valence band $E_V$ and the conduction band $E_C$ (figure 6, curves 1, 3). The positions of $E_D$ and $E_V$ (figure 6, curves 1, 2) in p-$Bi_2Te_3$ and the solid solutions with a small number of substituted atoms in the Bi sublattice obtained in the study are differ from [39], which caused by the features of the growth technology and also by the influence of excess Te. In the solid solutions at x> 1.5, the $E_D$ and $E_V$ positions are in agreement with [39].

The $E_g$ values (figure 7, curve 1) obtained from the differential conductance $dI_t/dU_t$ are larger than $E_g$ received from investigation of optical absorption edge, especially in $Bi_2Te_3$ [40, 43] and p-$Bi_{2-x}Sb_xTe_3$ [43], which is probably caused by the effect of a change in the density of states in the materials with inversion of the energy gap edges in narrow-band topological insulators [46].

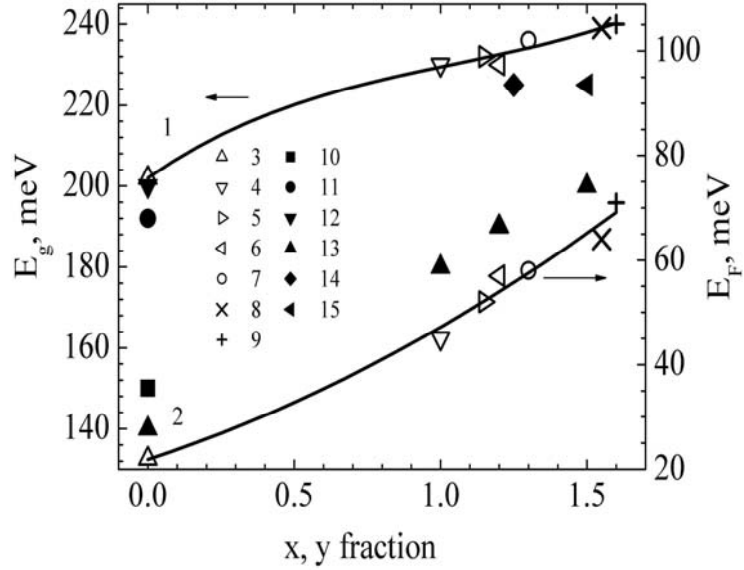

Figure 7. Position of the energy gap $E_g$ (1) and the Fermi Level $E_F$ (2) in dependence on the composition of p-$Bi_2Te_3$ and solid solutions.
p-$Bi_2Te_3$ – *3*, p-$Bi_{2-x}Sb_xTe_{3-y}Se_y$: *4* – (x= 1, y=0.06), *5* – (x= 1.15, y=0.06), *6* – (x= 1.2, y=0.09), *7* – (x= 1.3, y=0.09), *8*– (x= 1.55, y=0), *9* – (x= 1.6, y= 0.06), *10* – [40], *11* – [41], *12* – [42], *13* – [43], *14* – [44], *15* – [45].

The $E_g$ value obtained for p-$Bi_2Te_3$ by means of STS [42] is close to our data (figure 7). As shown in figure 7, the Fermi energy level with respect to the top of the valence band is located in the energy gap and slightly shifts depending on compositions of the solid solutions (figure 7, curve 2).

### 6. Surface states of Dirac fermions and thermoelectric properties

The position of the Dirac point $E_D$, depending on the composition of the solid solutions (figure 6, curve 1) together with the Fermi velocity values $v_F$ allow to estimate the surface state parameters of the Dirac fermions: the surface concentration $n_s$, the wave vector $k_F$. In these estimates, we used data for the Fermi velocity $v_F$ in the $Bi_{2-x}Sb_xTe_3$ solid solutions given in [39]. The surface concentration $n_s$ is defined as:

$$n_s = \frac{1}{4\pi} k_F^2 \quad (1),$$

where $k_F = \frac{|E_D|}{v_F}$

Figure 8 shows the dependences of the surface concentration of Dirac fermions $n_s$, the Fermi velocity $v_F$, and the wave vector $k_F$ in the p-$Bi_2Te_3$ and the p-$Bi_{2-x}Sb_xTe_{3-y}Se_y$ solid solutions for different composition, obtained in accordance with equation (1).

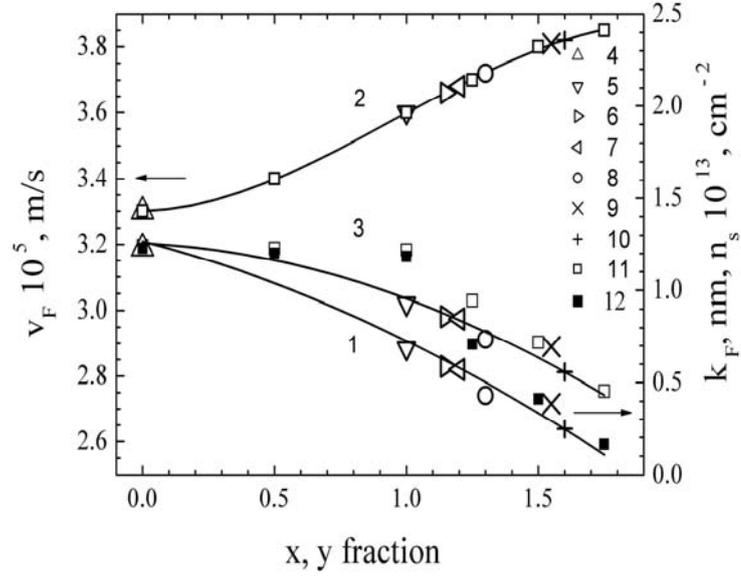

Figure 8. The surface concentration of Dirac fermions $n_s$ (1), the Fermi velocity $v_F$ (2), and the wave vector $k_F$ (3) in p-$Bi_2Te_3$ and solid solutions versus composition.
p-$Bi_2Te_3$ – *4*, p-$Bi_{2-x}Sb_xTe_{3-y}Se_y$: *5* – (x= 1, y=0.06), *6* – (x= 1.15, y=0.06), *7* – (x= 1.2, y=0.09), *8* – (x= 1.3, y=0.09), *9* – (x= 1.55, y=0), *10* – (x= 1.6, y=0.06), *11* – $k_F$, $v_F$, *12* – $n_s$ [39].

With increase of atomic substitutions in the solid solutions, the velocity $v_F$ enhances by approximately 20%, and the surface concentration of $n_s$ decreases from 1.25 to 0.25 $10^{13}$ cm$^{-2}$ in comparison with p-$Bi_2Te_3$ (figure 8, curves 1, 3). The solid solutions with large substitutions of atoms (x = 1.55 and x = 1.6, y = 0.06) exhibit different surface concentrations of about 0.38 and 0.25 $10^{13}$ cm$^{-2}$ with high values of velocity $v_F$ of about 3.8 $10^5$ m/s (figure 8, curve 3), which leads to an increase in the mobility of Dirac fermions in the surface layer in these compositions.

The results obtained in the study of the differential conductance $dI_t/dU_t$ were analyzed in conjunction with the thermoelectric properties of the materials based on bismuth telluride [3, 25, 28]. It allows us to establish correlation between the parameters of the topological surface states of the Dirac fermions and the basic thermoelectric properties of bulk thermoelectrics: the Seebeck coefficient S, the power factor $S^2\sigma$, and the material parameter proportional to the figure–of–merit Z (figure 9).

In the p-$Bi_{2-x}Sb_xTe_{3-y}Se_y$ solid solutions, the $S^2\sigma$ power factor and the $(m/m_0)^{3/2}\mu_0$ material parameter enhance with increase of the atomic substitutions (Table 1) induced by charge carrier scattering and the parameters of constant-energy ellipsoids in bulk materials [1, 7, 29-31]. Figure 9 shows that the power factor $S^2\sigma$ (curves 2) and the material parameter $(m/m_0)^{3/2}\mu_0$ (curves 3) enhance with shifting of the Dirac point $E_D$ to the top of the valence band that correlate with increasing of atomic substitutions in the solid solutions.

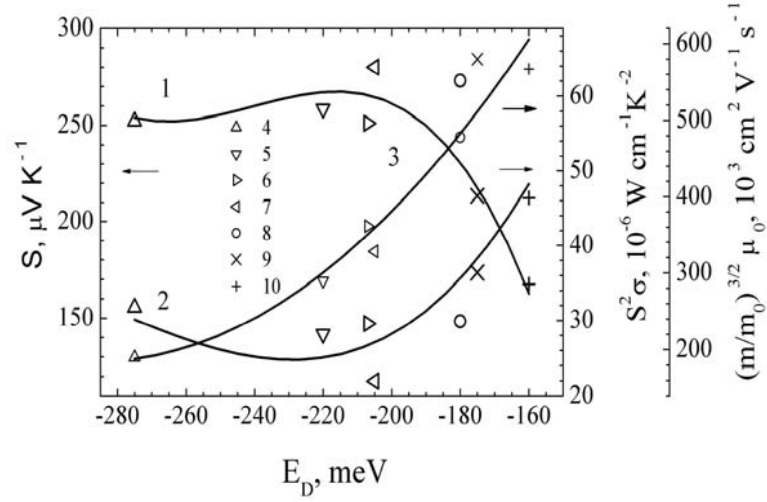

Figure 9. Dependence of the Seebeck coefficient S (1), the power factor $S^2\sigma$ (2) and the material parameter $(m/m_0)^{3/2}\mu_0$ (3) on the Dirac point energy $E_D$ in p-$Bi_2Te_3$ and solid solutions.
p-$Bi_2Te_3$ – *4*, p-$Bi_{2-x}Sb_xTe_{3-y}Se_y$: *5* – (x= 1, y=0.06), *6* – (x= 1.15, y=0.06), *7* – (x= 1.2, y=0.09), *8* – (x= 1.3, y=0.09), *9* – (x= 1.55, y=0), *10* – (x= 1.6, y=0.06).

The largest $E_D$ shift is observed in the compositions at x = 1.55 and x = 1.6, y = 0.06 with values of the Seebeck coefficient of about 170 μV K$^{-1}$ (Table 1). These solid solutions are efficient in the temperature range near room and at higher temperatures [28]. In the solid solutions with smaller substitutions of atoms (x = 1-1.3, y = 0.06 and 0.09), the $E_D$ shift decreases, and the values $S^2\sigma$ and $(m/m_0)^{3/2}\mu_0$ also reduce (figure 8, curves 2, 3). These compositions with high values of the Seebeck coefficient in the interval of 250-280 μV K$^{-1}$ (Table 1) are optimal in the temperature range below room temperature [1].

As shown in figure 9, in the solid solutions at x>1.5 with the highest parameters $S^2\sigma$ and $(m/m_0)^{3/2}\mu_0$, the surface fermion concentration $n_s$ decreases and the Fermi velocity $v_F$ increases, which leads to an enhance of the mobility in the surface layer as compared with the solid solutions at smaller substitutions of atoms. Unlike solid solutions with large substitutions of atoms for x> 1.5, the surface concentration of fermions $n_s$ increases in the compositions at x = (1 - 1.3), y = 0.06 and 0.09, which indicates enhance of the influence of the surface states of the Dirac fermions.

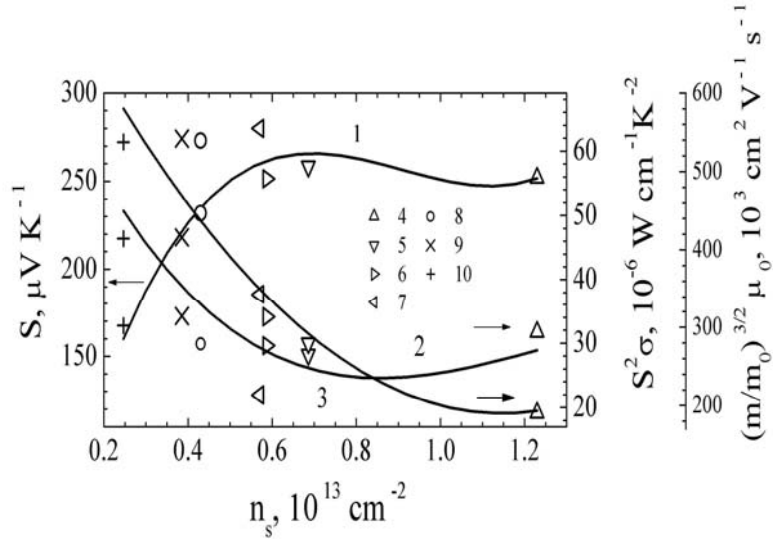

Figure 10. The Seebeck coefficient S (1), the power factor $S^2\sigma$ (2) and the material parameter $(m/m_0)^{3/2} \mu_0$ (3) in dependence on the surface Dirac fermion concentration $n_s$ in the p-$Bi_2Te_3$ and the p-$Bi_{2-x}Sb_xTe_{3-y}Se_y$ solid solutions. Point notations as in figure 9.

Thus, the growth in the power factor and the material parameter in the p-$Bi_{2-x}Sb_xTe_{3-y}Se_y$ solid solutions with increasing of atomic substitutions (x> 1.5) is accompanied by significant shift of the Dirac point $E_D$ to the top of the valence band and by an enhance in the Fermi velocity $v_F$, which leads to raise in charge carrier mobility in surface layer. In the compositions at x = (1 - 1.3) with a smaller shift of the Dirac point $E_D$, the influence of the surface states of the Dirac fermions is related to increase of the surface concentration $n_s$.

## 7. Conclusion

The morphology of the interlayer van der Waals surface (0001) and differential tunneling conductance were studied by STM/STS technique in the p-$Bi_2Te_3$ and p-$Bi_{2-x}Sb_xTe_{3-y}Se_y$ thermoelectrics. The surface atom arrangement corresponds to hexagonal close-packed structure. The height of longwave modulation along the profile of the interlayer surface is associated with local distortions in the density of surface electronic states under substitution of Bi → Sb, Te → Se atoms in solid solutions.

Pictures of the reciprocal lattice obtained by the FFT method were used to estimate the local lattice constants, which are qualitatively consistent with the results of x-ray studies. In the vicinity of the Γ point of the Brillouin zone, Friedel type oscillations were observed, which relate to interference of quasiparticle excitations of surface electrons on defects in topological insulators. The effect of Se atoms was also observed in the FFT pictures as enhanced of the second-order spectral components due to an increase of Se and excess Te in the compositions with substitutions of atoms in both Bi and Te sublattices.

From the analysis of the differential tunneling conductance, it was established that in the p-$Bi_{2-x}Sb_xTe_{3-y}Se_y$ solid solutions, the Dirac point $E_D$ moves to the top of the valence band in dependence on the compositions with an increase in substitutions of atoms in the Bi sublattice. Substitution of atoms in the Te sublattice leads to an additional $E_D$ shift even for small amounts of Se atoms. However, in all the studied materials, the Dirac point remains in the valence band.

The shift of the valence and conduction band edges increases with enhance of the atomic substitutions in the solid solutions, which leads to an increase in the energy gap $E_g$ in comparison with the values obtained from the optical data. Such shift is explained by a change in the density-of- states due to inversion of the band gap edges in topological insulators.

It is established that the power factor $S^2\sigma$ and the material parameter $(m/m_0)^{3/2}\mu_0$ enhance with the shift of the Dirac point $E_D$ to the top of the valence band with increasing atomic substitutions in the solid solutions. The largest growth in $S^2\sigma$ and $(m/m_0)^{3/2}\mu_0$ with $E_D$ shift is observed in the compositions at x = 1.55 and x = 1.6, y = 0.06, which are known as the effective thermoelectrics at room and higher temperatures. In these compositions, an increase in the Fermi velocity leads to enhance in the mobility of the charge carriers, which determines the contribution of the surface states to the thermoelectric properties. In the compositions with less atom substitutions at x = (1-1.3) and y = 0.06 and 0.09, which are effective thermoelectrics below room temperature, a similar contribution of surface states is related to increase in the concentration of Dirac fermions.

**Acknowledgements**

This study was partially supported by Russian Foundation for Basic Research Project No. 16-08-00478a. The authors are grateful to A N Klimov (Prokhorov Institute, Moscow) and V N Petrov (Ioffe Institute, St. Petersburg) for the fruitful help in STM spectroscopy adjustment.

**Highlights**

- Topological characteristics of the Dirac fermion surface states were determined in p-$Bi_{2-x}Sb_xTe_{3-y}Se_y$ solid solutions by STM/STS method.
- Dirac point and Fermi level positions were acquired.
- Correlation between the topological characteristics, the thermoelectric power factor and the material parameter was found.